# Single-layer $MoS_2$ mechanical resonators

Andres Castellanos-Gomez*, Ronald van Leeuwen, Michele Buscema, Herre S. J. van der Zant, Gary A. Steele and Warner J. Venstra.

Kavli Institute of Nanoscience, Delft University of Technology, Lorentzweg 1, 2628 CJ Delft, The Netherlands.

E-mail: a.castellanosgomez@tudelft.nl

Atomically thin two-dimensional crystals are prospective materials for nano-electromechanical systems due to their extraordinary mechanical properties [1-7] (high Young's modulus, elasticity and breaking strength) and low mass. Among this family of 2D materials, graphene is the most studied one so far. Graphene mechanical resonators [8] have been already employed as mass and pressure sensors [9] and provide a platform to study interesting nano-mechanical phenomena such as nonlinear damping [10]. Nevertheless, the lack of a bandgap in graphene may limits its usefulness in certain applications requiring a semiconducting material. Molybdenum disulfide ($MoS_2$), a semiconducting analogue to graphene [11-13], presents excellent mechanical properties similar to graphene [5, 6] in combination to a large intrinsic bandgap [14-17]. Although multilayered $MoS_2$ resonators have been recently fabricated by Lee *et al.* [18] (one device 9 layers thick, the remainder devices 20 to 100 layers thick), single layer $MoS_2$ mechanical resonators have not been demonstrated so far. Unlike multilayer $MoS_2$, which has an indirect bandgap of 1.2 eV, monolayer $MoS_2$ is a direct bandgap semiconductor (1.8 eV) with potential applications in photodetection [19-22], photovoltaics [23] and valleytronics [24-26]. Therefore, fabrication of single-layer $MoS_2$ mechanical resonators is a first and necessary step towards nano-electromechanical systems exploiting the direct bandgap of monolayer $MoS_2$. Here, we demonstrate the fabrication of single-layer $MoS_2$ mechanical resonators. The fabricated resonators have fundamental resonance frequencies in the order of 10 MHz to 30 MHz (depending on their geometry) and their quality factor is about ~55 at room temperature in vacuum. We find that the mechanics of these single-layer resonators lies on the membrane limit (tension dominated) while multilayered $MoS_2$ resonators can be modeled as circular plates (bending rigidity dominated). We also demonstrate clear signatures of nonlinear behaviour of the single-layer $MoS_2$ resonators, thus providing a starting point for future investigations on the nonlinear dynamics of $MoS_2$ nanomechanical resonators.

Single layer $MoS_2$ mechanical resonators have been fabricated by transferring atomically thin flakes of $MoS_2$ onto a $SiO_2$/Si substrate pre-patterned with holes. We have found that direct exfoliation of $MoS_2$ onto the pre-patterned substrates yield a low density of flakes, and no suspended single-layer $MoS_2$ devices could be fabricated using this method. We have therefore employed an all-dry transfer technique to deposit the $MoS_2$ flakes onto the pre-patterned substrates, similar to the one described in [27, 28] (see Experimental Section and Supporting Information for a more detailed description of the transfer method). Single-layer devices can be identified at glance by means of optical microscopy [29, 30]. Confirmation of the exact layer thickness is performed by combination of atomic force microscopy, Raman spectroscopy [31] and photoluminescence [14-16] (see Experimental section and Supporting Information for more details).

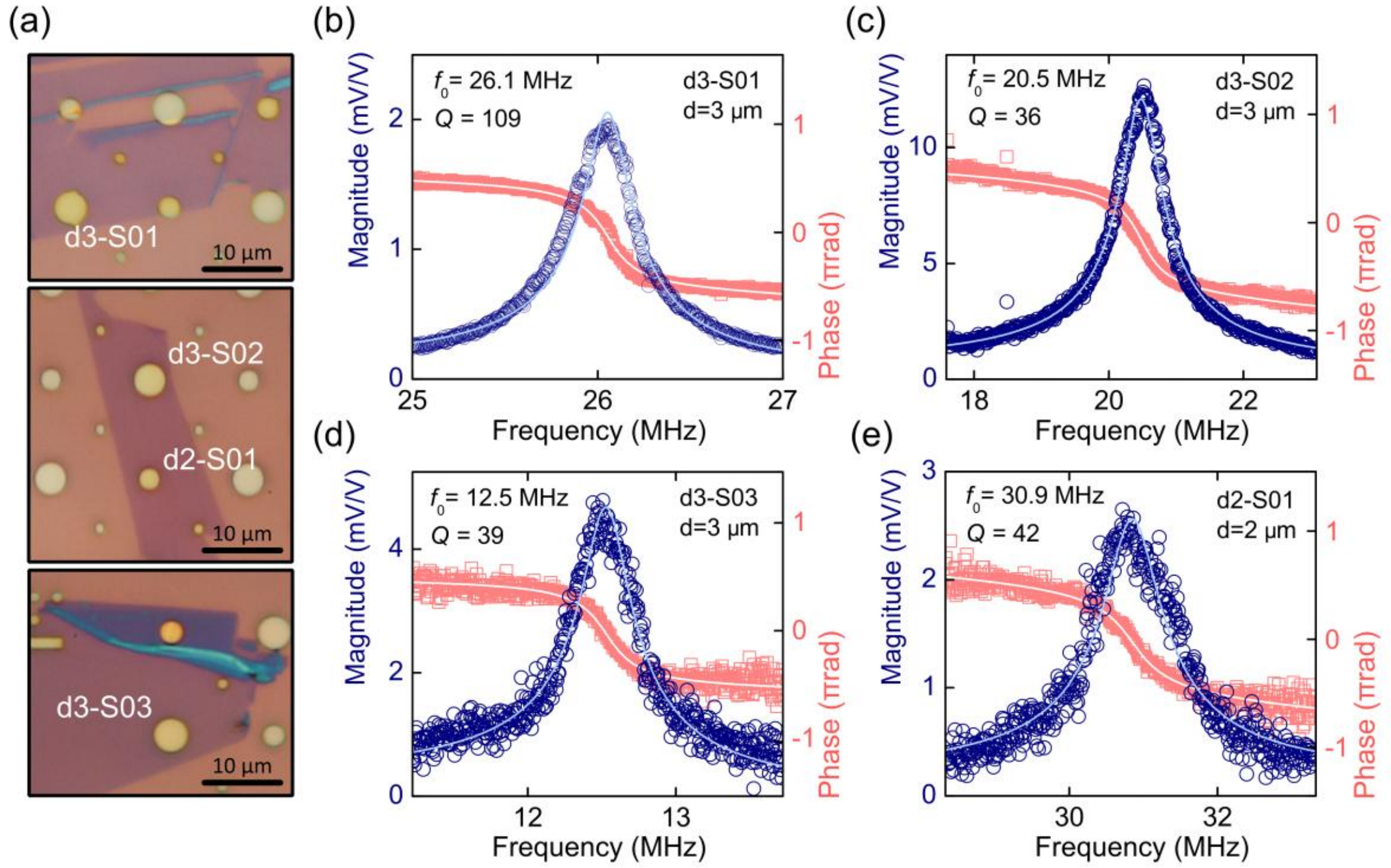

**Figure 1. Mechanical resonances of single-layer $MoS_2$.** (a) Optical microscopy image of single-layer $MoS_2$ flakes transferred onto a pre-patterned substrate with holes of different diameter. (b) to (e) Frequency response lines measured for different single-layer $MoS_2$ mechanical resonators. The experimental data (magnitude and phase) have been fitted to a damped-driven harmonic oscillator model (solid lines) to extract their resonance frequency and quality factor.

**Figure 1**a shows optical images of single layer $MoS_2$ mechanical resonators. The motion of the drum resonators is detected using an optical interferometer (see the Experimental section and the Supporting Information for more details about the experimental setup) [8]. Figure 1b-e shows the resonance spectra obtained for the single-

layer $MoS_2$ resonators highlighted in Figure 1a. The resonance frequency of these devices is in the order of 10 MHz to 30 MHz which is comparable to that of graphene resonators with similar geometries (16 MHz to 30 MHz)[32]. The quality factor of single-layer $MoS_2$ resonators have been obtained by fitting the magnitude and phase of the frequency response lines to a driven damped harmonic oscillator model (light lines in Figure 1b-e). The $Q$ factors of all the studied single-layer devices are between 17 and 105, with an average value of $Q$ = 54 ± 30, which is about a factor 3-4 lower than graphene drums with similar geometries [32] ($Q$ = 195 ± 15), and in the same order as those recently reported by Lee *et al*. for thicker $MoS_2$ resonators [18].

We have further characterized thicker $MoS_2$ resonators to determine whether the dynamics of $MoS_2$ resonators is dominated by their initial tension (membrane-limit) or by their bending rigidity (plate-limit). For a membrane-like resonator the fundamental frequency is given by [33]

$$f = \frac{2.4048}{\pi d}\sqrt{\frac{T}{\rho t}} \quad [1],$$

where $d$ is the resonator diameter, $T$ its initial pre-tension (in N/m), $\rho$ its mass density and $t$ the thickness. For a plate-like circular resonator, on the other hand, the frequency is given by [33]

$$f = \frac{10.21}{\pi}\sqrt{\frac{E}{3\rho(1-\nu^2)}}\frac{t}{d^2} \quad [2],$$

where $E$ is the Young's modulus and $\nu$ is the Poisson's ratio ($\nu$ = 0.125 [34]). Therefore one would expect a clearly different thickness dependence of the resonance frequency for membrane-like resonators ($f \propto t^{-1/2}$) and plate-like resonators ($f \propto t$). This indicates that thin resonators should behave as membranes (dominated by their initial pre-tension) while thick resonators should behave as circular plates. A cross-over from the membrane-to-plate behavior is also expected [6]. The resonance frequency of the mechanical resonators in that cross-over regime can be calculated as

$$f = \sqrt{f_{membrane}^2 + f_{plate}^2} \quad [3],$$

where $f_{\mathrm{membrane}}$ and $f_{\mathrm{plate}}$ are calculated according to [1] and [2] respectively.

**Figure 2**a shows the measured resonance frequency as a function of the thickness for several $MoS_2$ resonators (with a diameter of 2 μm and 3 μm). A total number of 94 devices with thicknesses ranging from single-layer to ~95 layers have been studied. The complete dataset is presented in the Supporting Information. We find a noticeable dispersion of the frequency (and the quality factor, see Supp. Info.) for devices with analogous geometry. Similar dispersion in resonance frequency and quality factors has been found for graphene mechanical resonators [35, 36] and it is mainly attributed to device-to-device variation of the pre-strain and clamping conditions (such as non-isotropic pre-tension). Another possible source of this dispersion could be the presence of non-uniformly distributed adsorbates on the resonators, the devices were fabricated by an all-dry stamping method and the measurements have been carried out in vacuum which minimizes the presence of fabrication residues or atmospheric adsorbates.

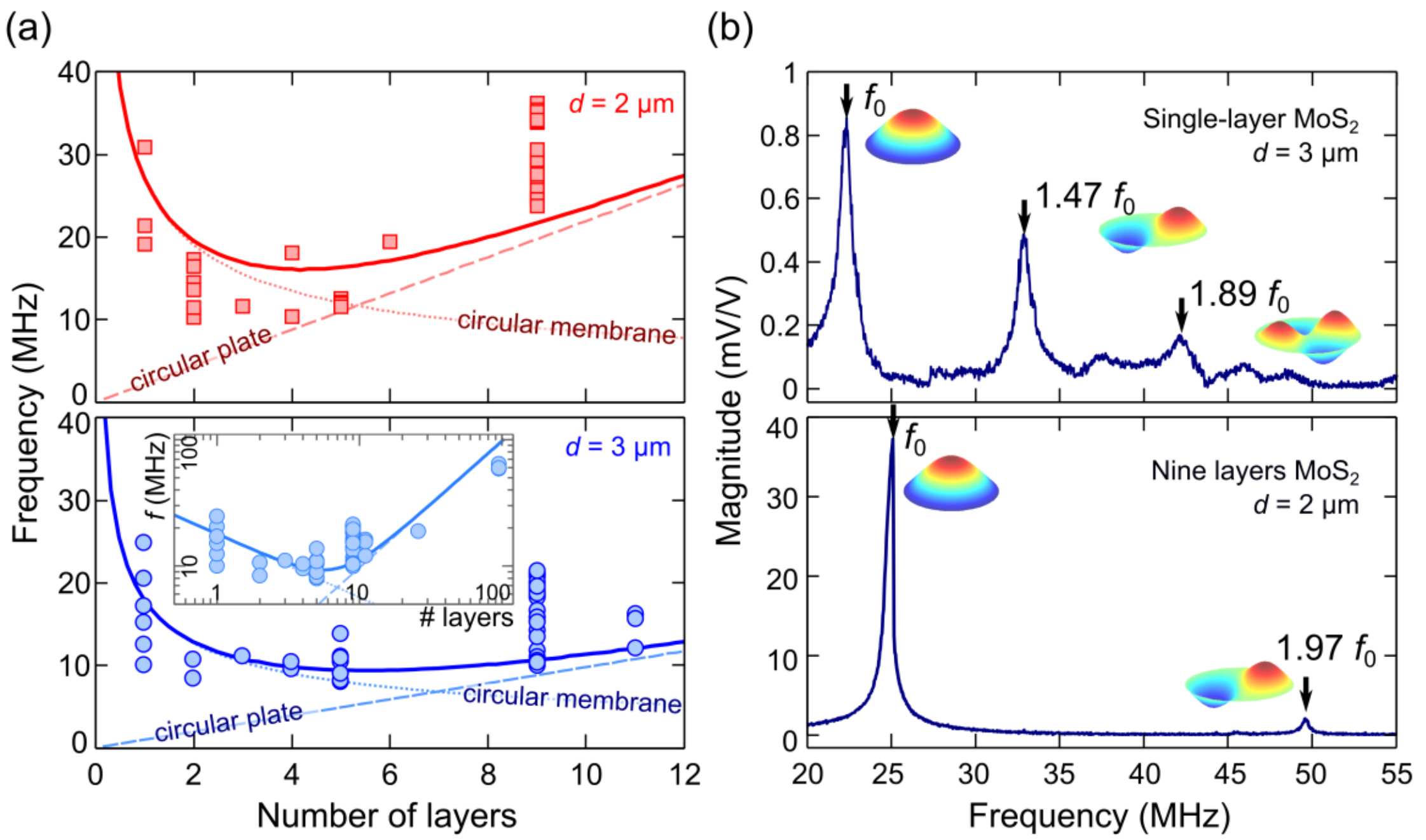

**Figure 2. Membrane to plate crossover in $MoS_2$ mechanical resonators.** (a) Measured resonance frequency as a function of the number of layers for $MoS_2$ mechanical resonators 2 μm in diameter (red squares in the top panel) and 3 μm in diameter (blue circles in the lower panel). The lines indicate the calculated frequency *vs.* thickness relationship for different cases: circular membrane (dotted), circular plate (dashed) and the combination of these two cases (solid). A The inset in the lower panel in (a) includes data measured for seven thick $MoS_2$ resonators (more than 30 layers), which are in the plate limit. (b) Mechanical resonance spectra measured for a single-layer $MoS_2$ resonator (top panel) and a nine layers thick $MoS_2$ resonator (lower panel), showing higher order modes. While for the single-layer the higher modes appear at frequencies similar to those expected for a circular membrane, for the thicker resonator the resonance spectra agrees with a circular plate.

The calculated resonance frequencies in the membrane and plate limit, as well as in the cross-over regime are plotted as dotted, dashed and solid lines respectively. A Young's modulus of $E$ = 300 GPa and a pre-tension $T$ = 0.015 N/m have been used for the calculation of the resonance frequencies, using Expressions [1] to [3], in agreement with the values found by nanoindentation experiments on drums fabricated using the same technique.[6, 37] The pre-tension value depends on the fabrication technique and it may be tuned by modifying the $MoS_2$ transfer process. Note that the calculated frequency *vs.* thickness relationship is not a fit to the experimental values but a guide to the eye to point out the different thickness dependences expected for membranes and plates. From the analysis of the frequency vs. thickness relationship (Figure 2a) one can see how the membrane term in Eq. [3] can be neglected for flakes thicker than 10-12 layers while the plate term can only be neglected for devices thinner than 3-4 layers. $MoS_2$ flakes 4 to 10 layers thick are in the cross-over regime where membrane and plate terms in Eq. [3] are comparable. The inset in Figure 2a (lower panel) includes measured resonance frequencies for much thicker $MoS_2$ devices, which follow the trend expected for plate-like resonators. This membrane-to-plate crossover is in good agreement with previous works based on nanoindentation in freely suspended $MoS_2$ layers.[6, 37] While for flakes thicker than 10 layers the force *vs*. deformation relationship could be modeled as elastic plates, flakes 5 to 10 layers in thick required a mechanical model including both membrane and plate terms.

A further proof of the membrane-to-plate cross-over is given by the analysis of higher eigenmodes. Figure 2b shows the broadband resonance spectrum measured on a 3 µm diameter single-layer $MoS_2$ resonator. Two higher eigenmodes at 1.47 $f_0$ and 1.89 $f_0$ are found, which are in reasonable agreement with the expected higher eigenmodes for a circular membrane (1.59 $f_0$ and 2.14 $f_0$). The matching between the measured values and the expected values is similar to that found for graphene drums [32] where the slight discrepancies with the predicted values were also attributed to non-uniform strain profile or to atmospheric adsorbates on the device. Note that for a circular plate, no eigenmodes around 1.59 $f_0$ are expected and the second eigenmode is expected to be around 2.08 $f_0$ [33]. Figure 2b (lower panel) shows the resonance spectra for a 2 µm diameter 9 layer thick $MoS_2$ device which shows one higher eigenmode at 1.97 $f_0$ (in good agreement to the expected value for a circular plate, 2.08 $f_0$ [33]). We therefore conclude that while thick $MoS_2$ resonators (9 or more layers) behave as plates, single-layer $MoS_2$ resonators are in the membrane-limit (*i.e.* their dynamics is dominated by their initial pre-tension).

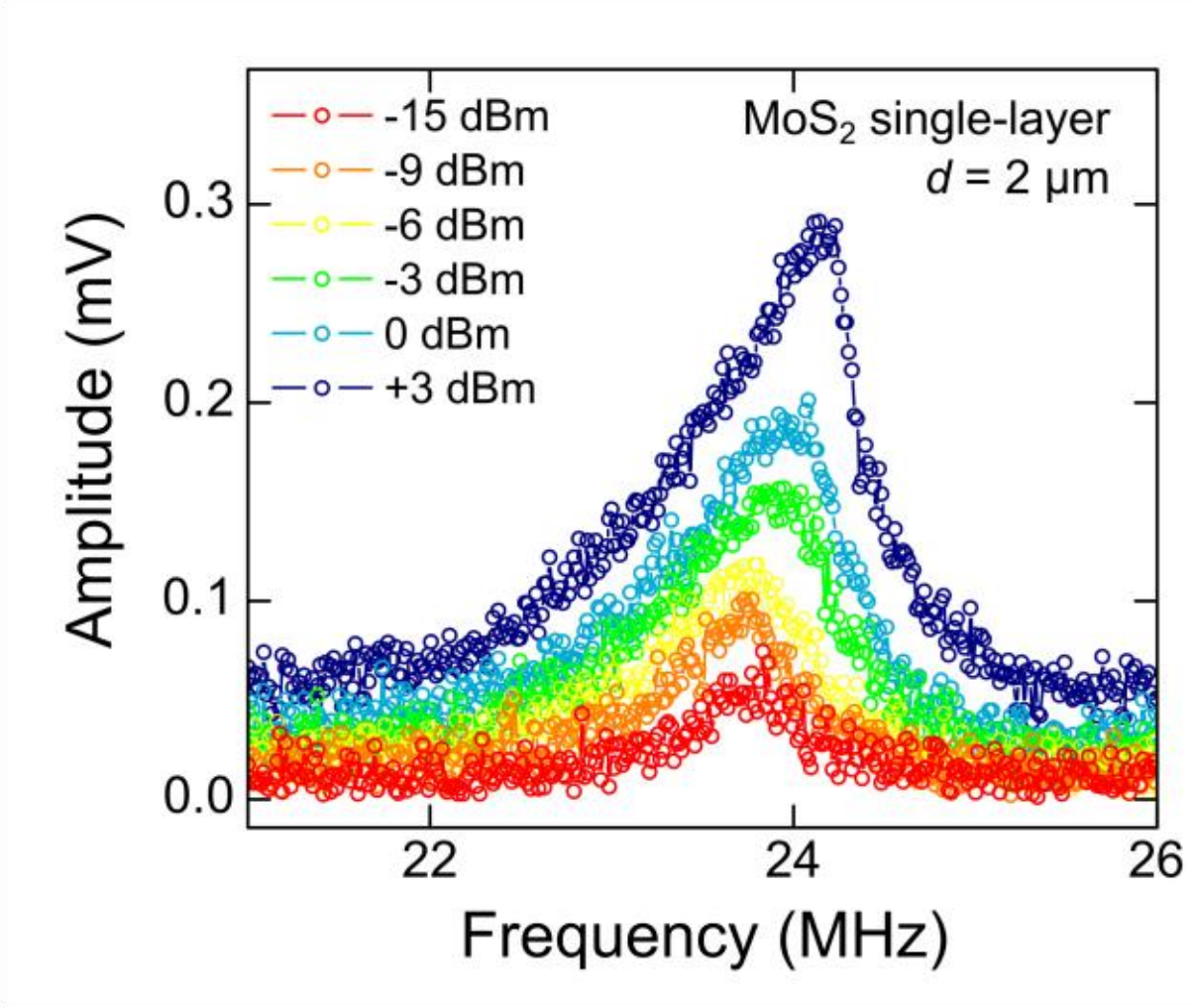

**Figure 3. Nonlinear resonance of a single-layer $MoS_2$ resonator.** Frequency response lines measured for a 2 µm diameter single-layer $MoS_2$ resonator at increasing optical excitation power. For driving powers higher than -9 dBm a clear deviation from a harmonic response is observed, indicating that the resonator is oscillating in the nonlinear regime. We estimate the membrane displacement to be on the order of 0.2 nm,[18] when clear signatures of a nonlinear response occur (blue response line, driven at + 3 dBm).

The photothermal excitation technique also enables driving the resonators beyond the regime of harmonic oscillations. **Figure 3** shows the vibration amplitude for a 2 µm single-layer $MoS_2$ resonator when driven through resonance at increasing optical power. For low amplitudes, up to -9 dBm, the spring constant of the membrane-like device is fully determined by the residual tension, which is independent of the displacement and gives rise to harmonic oscillations as observed in Figure 1. A clear deviation from a harmonic response is observed when the driving power is further increased, as the resonance frequency increases with the driving force and the response becomes strongly asymmetric. This is a signature of nonlinear behaviour, which is characteristic for nanoscale mechanical resonators when the vibration amplitude becomes comparable to the device thickness. In this regime the $MoS_2$ membrane stretches significantly, which induces an alternating tension. This displacement-induced tension provides an additional restoring force, which is cubic in the displacement and causes the observed stiffening of the frequency response. The internal coupling between the different vibration modes of the membrane, which arises from the same nonlinearity [38], opens new routes to tuning of the membrane $Q$-factor and the resonance frequency [39, 40]. New applications emerge for nonlinear $MoS_2$ membranes, such as sensitive detectors demonstrated in top-down fabricated nonlinear nanomechanical resonators[41], since the nonlinearity scales with the aspect ratio, these ultrathin membranes provide the ultimate devices to study the fundamental aspects of nonlinear nanomechanics.

In summary, we have demonstrated the fabrication of single-layer $MoS_2$ mechanical resonators with resonance frequencies in the order of 10-30 MHz and quality factors of about ~55 at room temperature in vacuum. These monolayer $MoS_2$ resonators are in the membrane limit (*i.e.*, tension dominated), in contrast to their thicker counterparts, which behave as circular plates. We also demonstrate clear signatures of nonlinear behavior of our single-layer $MoS_2$ resonators, thus providing a starting point for future investigations on the nonlinear dynamics of monolayer and few-layer $MoS_2$ nanomechanical resonators. Single-layer $MoS_2$ mechanical resonators combine excellent mechanical properties with an intrinsic large direct bandgap which makes them of great potential for application as nano-electromechanical systems. Moreover, since the bandstructure of single-layer $MoS_2$ can be modified by mechanical strain [42-48], new concepts of nanomechanical devices with mechanically tunable optoelectronic properties will emerge.

## *Experimental*

*Fabrication of $MoS_2$ mechanical resonators:* $MoS_2$ mechanical resonators have been fabricated by transferring $MoS_2$ layers with an all-dry transfer technique as described in Ref. [27, 28]. $MoS_2$ nanosheets have been deposited on a viscoelastic stamp (GelFilm® by GelPak) by mechanical exfoliation of natural $MoS_2$ (SPI Supplies, 429ML-AB) with blue Nitto tape (Nitto Denko Co., SPV 224P). The $MoS_2$ flakes are then transfer onto a pre-patterned substrate with holes of different diameters by pressing the viscoelastic substrate surface containing $MoS_2$ flakes against the pre-patterned substrate with the help of a micromanipulator. The viscoelastic substrate is then peeled-off very slowly using the micromanipulator.

*Optical identification:* To ensure optical visibility of ultrathin $MoS_2$ layers, the pre-patterned substrates are made from Si/$SiO_2$ wafers with a 285 nm thick thermally grown $SiO_2$ layer [29]. Single- and few-layer $MoS_2$ sheets are located under an optical microscope (Olympus BX 51 supplemented with a Canon EOS 600D camera) and their number of layers is estimated by their optical contrast [29].

*Thickness determination:* Atomic force microscopy, Raman spectroscopy and photoluminescence were also performed for the accurate determination of the number of layers. An atomic force microscope (Digital Instruments D3100 AFM with standard cantilevers with spring constant of 40 N/m and tip curvature <10 nm) operated in the amplitude modulation mode has been used to study the topography and to determine the number of layers of $MoS_2$ flakes previously selected by their optical contrast. A Renishaw *in via* system was used to perform micro-Raman spectrometry and photoluminescence measurements. The system was used in a backscattering configuration excited with visible laser light ($\lambda$ = 514 nm) to accurately determine the number of layers of the studied $MoS_2$ flakes [31, 49]. The spectra were collected through a 100× objective (NA = 0.95) and recorded with 1800 lines/mm grating providing the spectral resolution of ~ 0.5 $cm^{-1}$. To avoid laser-induced modification or ablation of the samples, all spectra were recorded at low power levels P ~ 250 µW [50].

*Interferometric motion detection:* A Helium-Neon laser ($\lambda$ = 632.8 nm) is focused on the suspended part of the flake, which acts as the moving mirror of an optical interferometer. The silicon surface at the bottom of the drum acts as the fixed mirror. The motion of the $MoS_2$ resonator changes the cavity length, which causes constructive or destructive interference. Thus, the motion of the resonator modulates the intensity of the light emitted from the cavity, which is detected using a photodiode. We employ a blue diode laser ($\lambda$ = 405 nm) with an optical output below 1 mW for photothermal excitation of the resonators, which is efficient given the small thermal time constants at the nanoscale. The measurements are carried out in vacuum (~$10^{-5}$ mbar) to avoid viscous damping from ambient gas molecules.

*Acknowledgements*

The authors would like to acknowledge Peter G. Steeneken for useful discussions. A.C-G. acknowledges financial support through the FP7-Marie Curie Project PIEF-GA-2011-300802 (‘STRENGTHNANO’). R.v.L. acknowledges financial support from NanoNextNL, a micro and nanotechnology consortium of the Government of the Netherlands and 130 partners. M.B. acknowledges the financial support of the Dutch organization for Fundamental Research on Matter (FOM). W.J.V. acknowledges funding from the European Union's Seventh Framework Programme (FP7/2007-2013) under Grant Agreement n° 318287, project LANDAUER.

# Supporting information:

# Single-layer $MoS_2$ mechanical resonators

Andres Castellanos-Gomez *, Ronald van Leeuwen, Michele Buscema,
Herre S. J. van der Zant, Gary A. Steele *and* Warner J. Venstra.

Kavli Institute of Nanoscience, Delft University of Technology, Lorentzweg 1, 2628 CJ Delft, The Netherlands.

*E-mail: a.castellanosgomez@tudelft.nl

## Fabrication of $MoS_2$ resonators

Freely suspended multilayered $MoS_2$ layers can be fabricated by directly exfoliating $MoS_2$ onto a pre-patterned substrate with holes or trenches. Although this method have been successfully employed by different groups to fabricate freely suspended single-layer graphene [1, 2], it typically yields a very low density of suspended thin $MoS_2$ flakes [3-5]. Moreover the fabrication of freely suspended single-layer $MoS_2$ has proven to be elusive using this method.

In order to overcome this limitation, here we employ an all-dry transfer technique based on elastomeric stamps. The details of this technique can be found in Ref. [6, 7]. Briefly:

1. $MoS_2$ flakes have been deposited on an elastomeric stamp (GelFilm® by GelPak) by mechanical exfoliation of natural $MoS_2$ (SPI Supplies, 429ML-AB) with blue Nitto tape (Nitto Denko Co., SPV 224P).
2. Atomically thin $MoS_2$ flakes are identified on the surface of the elastomeric stamp by transmission mode optical microscopy. The optical absorption of a single-layer $MoS_2$ is about a 5.5% (See Supp. Info. in Ref. [8]).
3. The stamp surface containing the $MoS_2$ flakes is mounted in a micromanipulator facing the pre-pattern substrate.
4. The stamp is brought into contact with the pre-patterned substrate by lowering the manipulator.
5. The stamp is peeled off very slowly using the micromanipulator.

As the viscoelastic stamps are transparent, this process can be monitored under an optical microscope (see **Figure S1**).

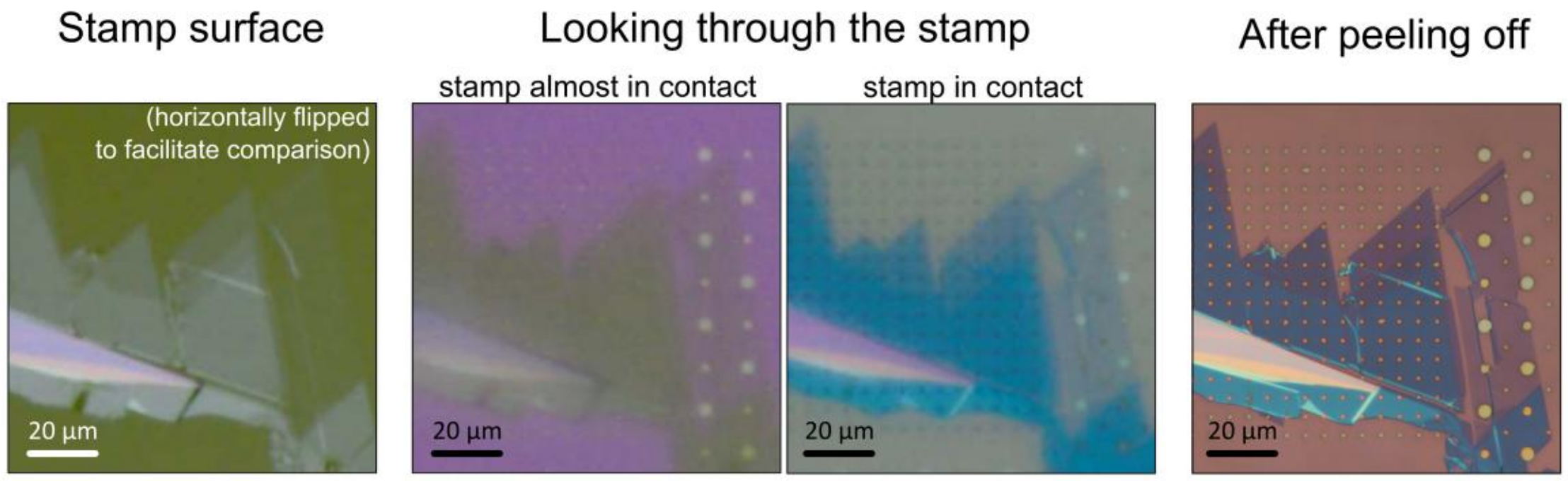

**Figure S1. Fabrication of $MoS_2$ resonators by an all-dry stamping method.** Optical microscopy images of the process to fabricate $MoS_2$ resonators. First $MoS_2$ flakes are transferred onto a viscoelastic stamp.

## Optical microscopy characterization

In order to ensure optical visibility of the transferred ultrathin $MoS_2$ layers, the pre-patterned substrates are made with Si/$SiO_2$ wafers with a 285 nm thick thermally grown $SiO_2$ layer [9]. Single- and few-layer $MoS_2$ sheets can be easily distinguished under an optical microscope (Olympus BX 51 supplemented with a Canon EOS 600D digital camera) because of their different optical contrast [9, 10]. **Figure S2** shows optical microscopy images of $MoS_2$ flakes with different number of layers.

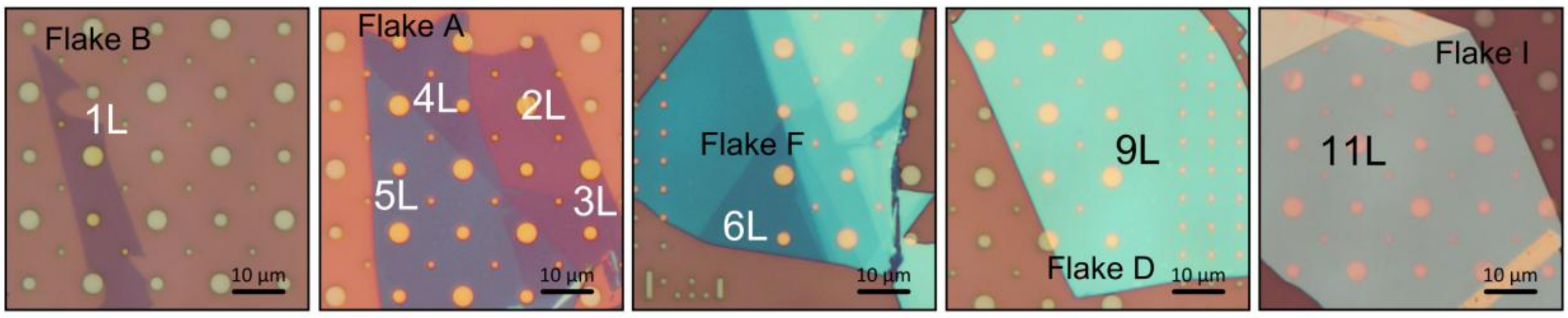

**Figure S2. Optical images of fabricated $MoS_2$ resonators with different thicknesses.** Optical microscopy images of the $MoS_2$ flakes transferred onto the pre-patterned substrate with holes. The apparent color of the flakes can be used to estimate at glance the number of layers.

## Atomic force microscopy characterization

Atomic force microscopy was also performed for a more accurate determination of the number of layers. An atomic force microscope (Digital Instruments D3100 AFM with standard cantilevers with spring constant of 40 N/m and tip curvature <10 nm) operated in the amplitude modulation mode has been used to study the topography and to determine the number of layers of $MoS_2$ flakes previously selected by their optical contrast. The topography of the studied $MoS_2$ resonators is also studied to discard resonators that have been wrinkled/broken during the fabrication procedure.

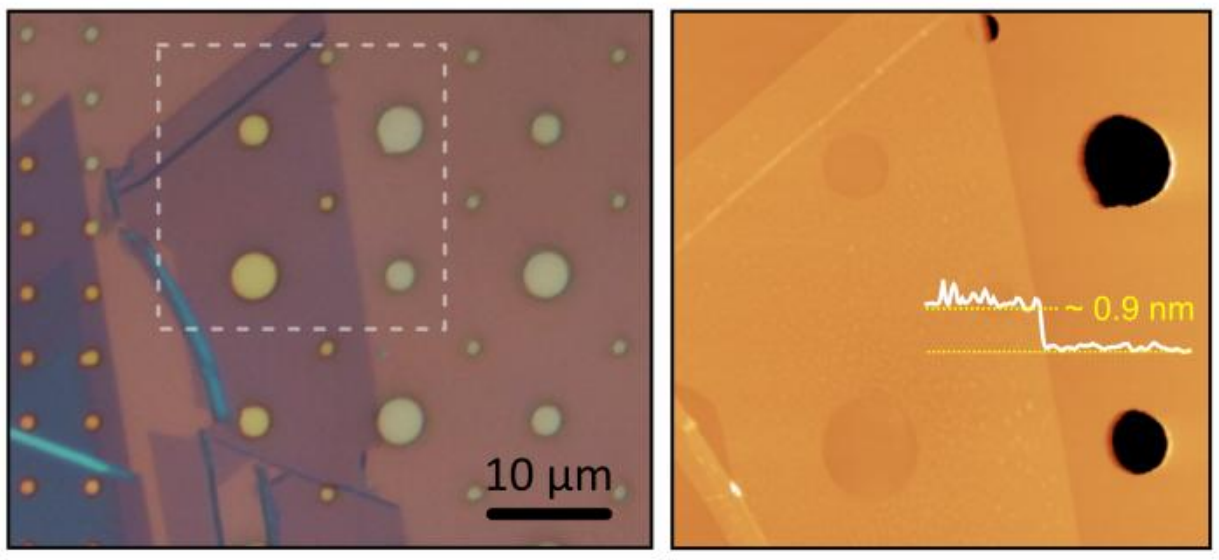

**Figure S3. Atomic force microscopy characterization of $MoS_2$ based resonators.** Atomic force microscopy image (right) of one single layer $MoS_2$ flake, previously selected because of its optical contrast (left). The thickness of the flake is in agreement with the values typically reported for single-layer $MoS_2$ on $SiO_2$/Si substrates.

## Raman and Photoluminescence Characterization

In this section, we present the characterization of the measured $MoS_2$ resonators via Raman and Photoluminescence (PL). It is well known that both the Raman and PL spectra have features that strongly depend on the number of $MoS_2$ layers. [11, 12] Therefore, we performed Raman and PL measurements to reliably assign the number of layers in each resonators. Measurements are performed in a micro-Raman spectrometer (*Renishaw in via*) in backscattering configuration with 100x magnification objective (NA=0.95) delivering a diffraction limited spot (~ 400 nm diameter). Excitation is provided by an Argon laser ($h\nu$ = 2.41 eV) with a typical power of ~ 250 µW to prevent spurious heating effects [13] and laser-induced ablation [14]. The elastic scattering is removed via a combination of a 50/50 beamsplitter and two notchfilters. The slits of the single-pass spectrometer (1800 lines/mm grating) are set to ~ 25 µm and the dispersed signal is collected by a CCD-array cooled by a Peltier element. The resolution is in the order of 0.5 $cm^{-1}$. Typical spectra span from 517 nm to 900 nm and it is therefore possible to record both the Raman and the PL part of the emission from $MoS_2$ simultaneously. This restricts the integration times to about 10 seconds. The measurements are performed on the supported part of the flakes to facilitate the comparison to previously published results [11, 15, 16].

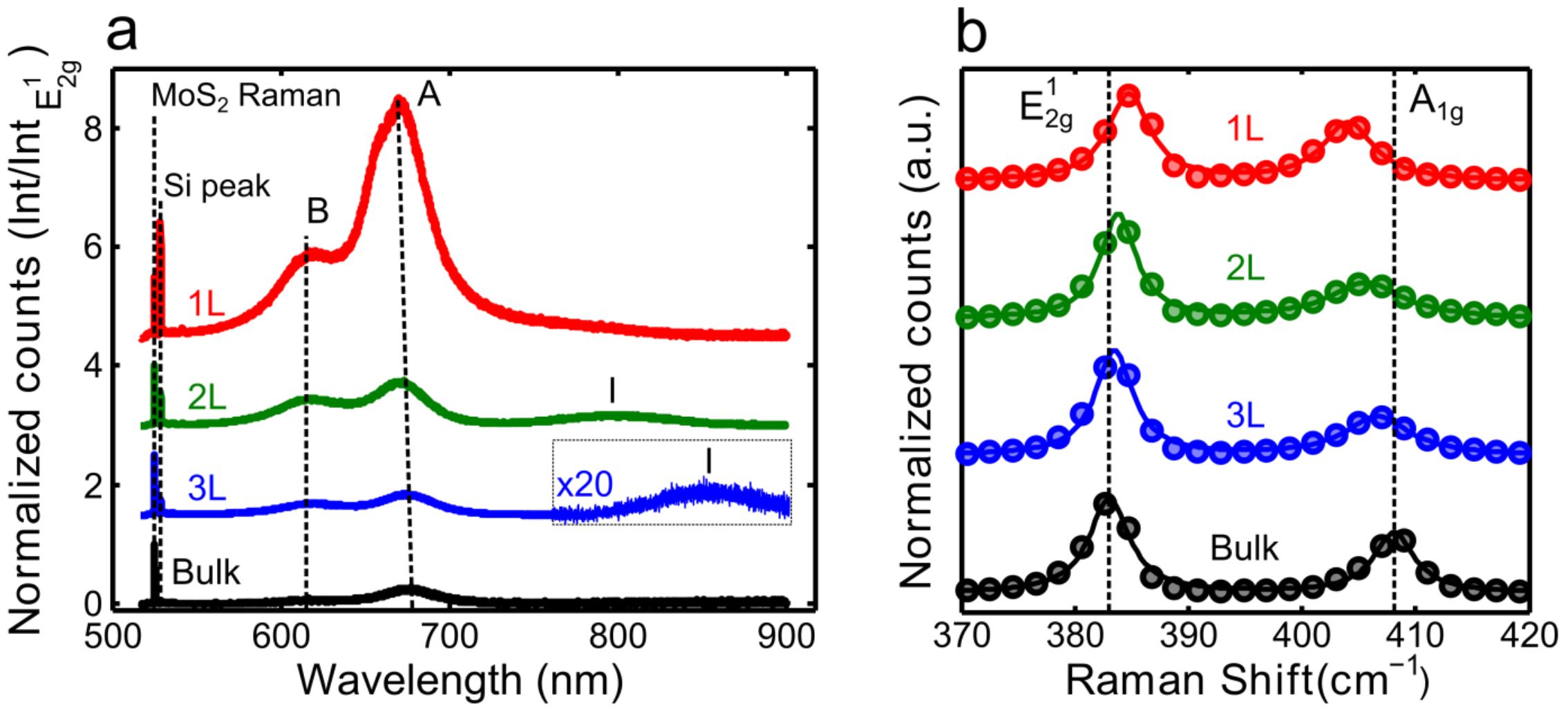

**Figure S4. Raman and photoluminescence characterization of the fabricated devices.** (a) Normalized emission intensity as a function of wavelength for 1 layer (red line), 2 layers (green line), 3 layers (blue line) – note the 20 times multiplication factor between 750 nm and 900 nm – and bulk (black line). The intensity is normalized to the intensity of the $E^1_{2g}$ Raman mode. The dashed lines indicate (from left to right) the position of the $MoS_2$ Raman peaks, the Si Raman peak, the B and A direct excitonic transitions. The label “I” indicates the indirect transition in the $MoS_2$ bandstructure. (b) Zoom in the Raman part of the spectra presented in panel a, with the same colour code. The dots are the experimental points while the lines are Lorentzian fit to the data. The dashed lines are the position of the $E^1_{2g}$ and $A_{1g}$ mode for bulk and ease the comparison between data for different number of layers. Note that in both panels the data are normalized to the intensity of the $E^1_{2g}$ mode and shifted vertically for clarity.

**Figure S4**a plots a selection of the full spectra as a function of the number of $MoS_2$ layers and Figure S4b shows a zoom-in in the Raman part of the spectra: the dots are the experimental data, the lines are Lorentzian fits. In both cases, the spectra are aligned to the Si Raman peak ( ~ 520 $cm^{-1}$ = 528.64 nm) and normalized to the intensity of the $E^1_{2g}$ Raman-active mode of $MoS_2$. This normalization procedure allows for direct comparison of the spectra through removal of systematic errors in the measurements (e.g. a slight variation in the excitation power) and accounting for the different laser absorption for flakes with different thicknesses. In Figure S4a the typical features of the PL of $MoS_2$ are present: the emission from the direct gap (known as A exciton, labelled with A in Figure S4a ) at about 670 nm, the emission from the valence-split exciton (known as B exciton, labelled as B in Figure S4a) at around 610 nm and the emission from the indirect transition (I) at about 800 nm for the bilayer. It can be seen that the A transition slightly redshift with increasing number of layers and drastically decreases in intensity (see **Figure S5**b). Moreover, the indirect transition (I) is absent in 1L MoS2 and appears from 2 layers and rapidly reduces in intensity and redshifts for higher number of layers.

These features are consistent with the assignment of the number of layers presented in Figure S4a.
Another proof of the assignment of the number of layers can be found in the Raman part of the spectra. Figure S4b plots a zoom-in in the Raman part of the spectra of Figure S4a. The two most prominent Raman-active modes are clearly visible: the lower wavenumber in-plane shearing mode ($E^1_{2g}$) and the higher wavenumber out-of-plane mode ($A_{1g}$). It can be clearly seen that the frequency of these two mode changes steadily with the number of layers: from bulk to 1 layer the $E^1_{2g}$ softens and the $A_{1g}$ hardens. This indicates that it is possible to use the difference between these modes to assign the number of layers (see figure S5a), in agreement with previous literature.[11]

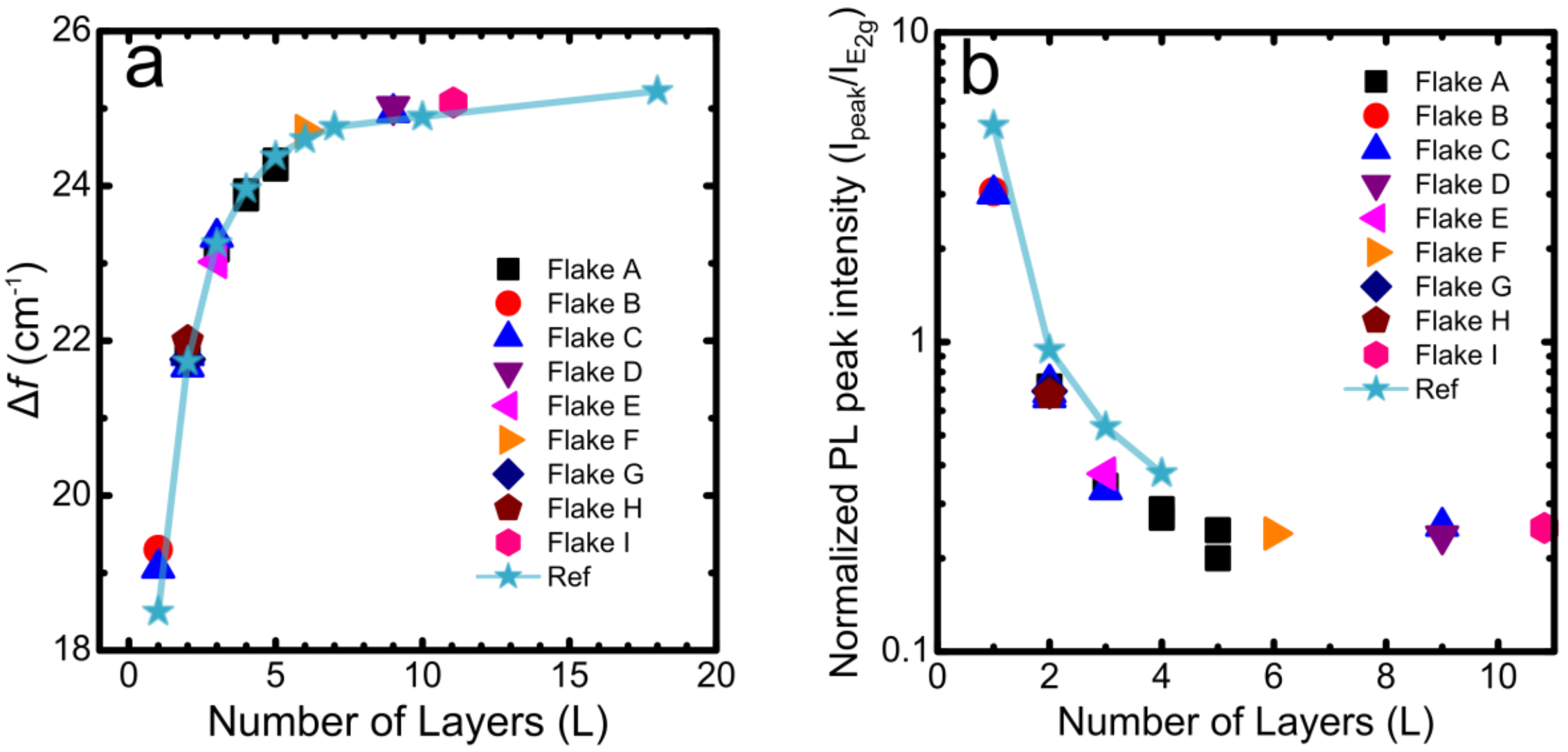

**Figure S5. Determination of the number of layers by Raman and photoluminescence.** (a) Frequency difference between the $A_{1g}$ and the $E^1_{2g}$ modes ($\Delta f$) as a function of the number of $MoS_2$ layers for this work (flakes from A to I) and from a reference sample. Measurements are performed on regions with different number of layers on the same flake. The frequencies of the modes are extracted from the Lorentzian fits shown in Figure S4b. (b) Semilogarithmic plot of the normalized PL peak emission *vs.* the number of $MoS_2$ layers for this work (flakes from A to I) and from a reference sample. The intensity is normalized to the $E^1_{2g}$ Raman mode intensity.

Figure S5a shows the difference between the $A_{1g}$ and the $E^1_{2g}$ modes ($\Delta f$) as a function of the number of $MoS_2$ layers for this work (flakes from A to I) and from a reference sample. The optical images of flakes A, B, D, F and I can be found in Figure S1. The reference sample was fabricate using the standard Scotch tape exfoliation method as previously reported.[17] It is possible to see the good agreement between the measurement performed on various zones in the flakes used in the present work and the reference samples. This confirms the robustness of the procedure to assign the number of layers.
Figure S5b plots the normalized intensity of the PL peak (~ A exciton emission) as a function of the number of $MoS_2$ layers for this work (flakes from A to I) and from the reference sample previously described. The PL emission intensity drastically drops with the number of layers and shows discrete values for different number of layers. There is fairly good agreement with the data measured in this work and the reference sample. We attribute the offset to different doping level of the fabricated devices with respect to the reference sample used.

For flakes thicker than 6 layers, Figure S5a and S5b saturate hampering the layer determination through Raman spectroscopy and photoluminescence. For flakes thicker than 6 layers, the thickness determination can be still done using AFM. Note that the uncertainty of the AFM is about a 5%.

## Optical interferometer setup

The motion of the suspended membrane is probed using a constant power linearly polarized HeNe laser (λ = 632.8 nm). The output power is adjusted with a neutral density filter, and the beam diameter is expanded using a 3× beam expander to match the aperture of the objective lens. A polarized beam splitter separates the incoming beam from the reflected light of the suspended drum. The entering beam is retarded by a quarter wave plate. Light emitted from the blue laser diode (*LP405-SF10*, λ = 405 nm), which is used for membrane excitation, is coupled in the main beam line using a dichroic mirror. The combined red and blue light passes through a semitransparent mirror and is focussed on the suspended membrane using a microscope objective lens (50×, NA=0.6). Reflected light from the membrane passes through the same objective lens and the semitransparent mirror, where part of the light is reflected to a CCD camera which is used for inspection and alignment. The blue light is reflected out of the returning light path using the dichroic mirror. The transmitted red light (λ = 632.8 nm) passes again the quarter wave plate and is directed towards a high speed photo detector (*NewFocus 1601*) using the polarizing beam splitter. The detector output is amplified (20dB, *Mini-Circuits ZFL1000LN*) and connected to the input port of a network analyzer (*Rohde & Schwartz, ZVB4*). Transmission measurements are carried out by modulating the diode laser, with typical optical modulation depth of 19 µW for the linear regime measurements (Figure 1 in the main text) and up to 142 µW to attain the nonlinear regime in monolayers (Figure 3 in the main text), powers as measured at the diode laser output.

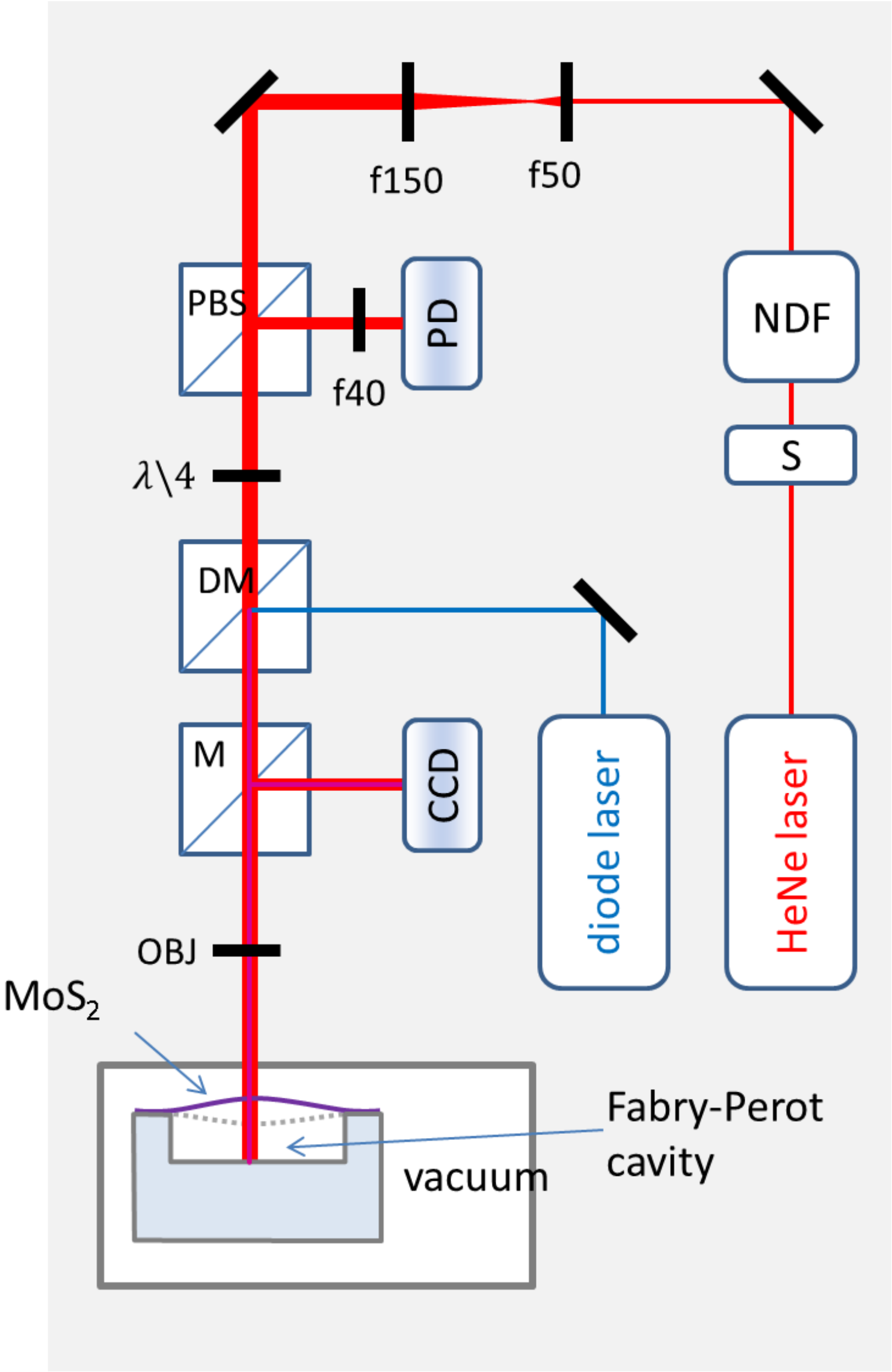

**Figure S6. Optical interferometer setup to measure the motion of $MoS_2$ based resonators.** S: shutter, NDF: neutral density filter, PBS: polarizing beam splitter, DM: dichroic mirror, M: 50/50 beamsplitter, PD: photodetector, CCD: camera, OBJ: 50× objective lens.

## Statistical analysis

**Figure S7** shows the histograms of the *Q* factor and *fQ* product values acquired for all the devices (94 devices) and for devices thinner than 10 layers (83 devices).

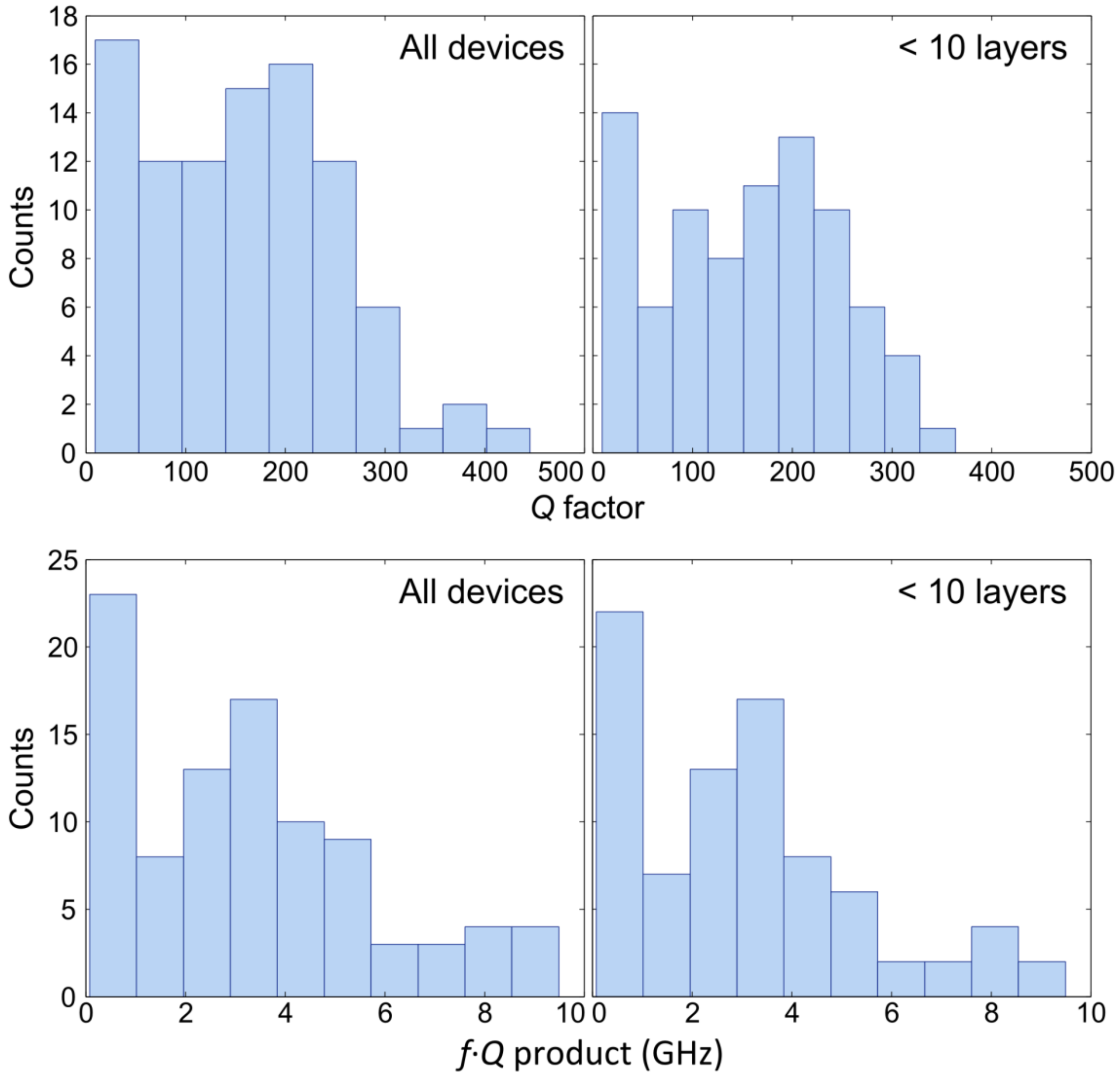

**Figure S7. Statistical analysis of the *Q* factor and the *fQ* product.** (top) Histograms of the *Q* factor values acquired for all the studied devices (left) and for devices thinner than 10 layers (right). (bottom) Histograms of the *fQ* product values acquired for all the studied devices (left) and for devices thinner than 10 layers (right).

## Thickness dependence of the resonance frequency

**Figure S8** shows the thickness dependence of the resonance frequency, displayed in Figure 2a of the main text (top and bottom panel as well as the inset). In Figure S8, it has been included the calculated frequency *vs.* thickness relationship using Expressions [1] to [3] of the main text with different Young's modulus (*E*) and pre-tension (*T*) values. The selected values are compatible with those previously obtained by means of AFM nanoindentation experiments on samples fabricated in a similar way [3, 4]. Note that this calculated relationship is not intended to be a fit to the experimental values but a guide to the eye to point out the different thickness dependences expected for membranes and plates.

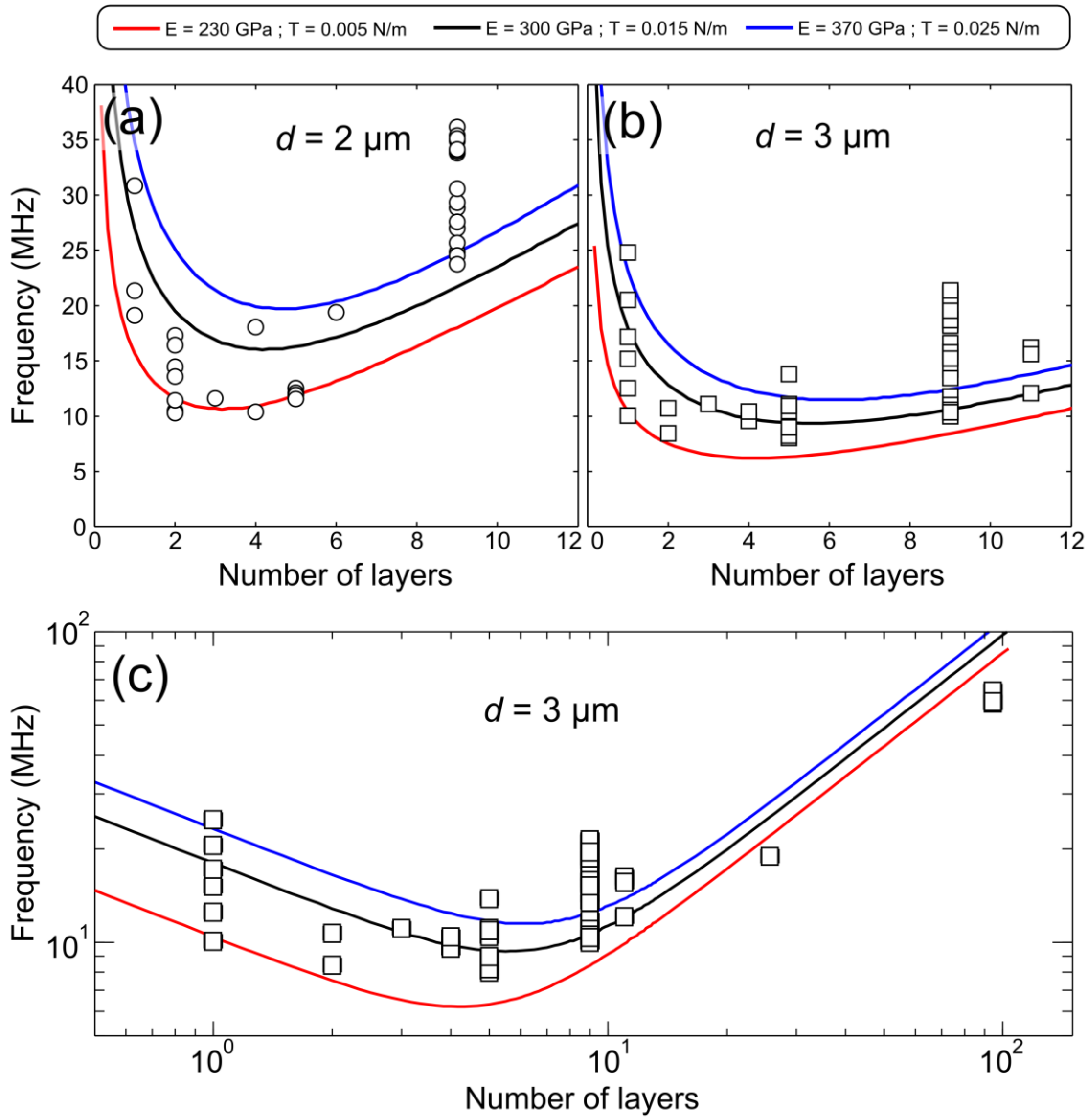

**Figure S8. Thickness dependence of the resonance frequency.** (a) Measurement of the thickness dependence of the resonance frequency for $MoS_2$ resonators 2 µm in diameter (symbols). The lines show the calculated relationship using expressions [1] to [3] of the main text using different Young's modulus (*E*) and pre-tension (*T*) values. (b) Same as (a) but for $MoS_2$ resonators 3 µm in diameter. (c) Shows the frequency vs. thickness relationship for the same devices shown in (b) but including 7 measurements for devices thicker than 11 layers.

## Complete datasets

**Table S1.** Complete dataset measured for $MoS_2$ resonators 3 μm in diameter. Note that the values marked with (*) have been determined only by AFM (see Raman and Photoluminescence Characterization section) and thus they present an uncertainty of about 5%.

| Sample | # layers | Frequency (MHz) | Quality factor | $A_{1g} – E^1_{2g}$ (cm$^{-1}$) | Norm. PL intensity |
|---|---|---|---|---|---|
| **d3-S01** | 1 | 26.10 | 109.00 | 19.02 | 3.43 |
| **d3-S02** | 1 | 20.50 | 35.75 | 19.30 | 3.06 |
| **d3-S03** | 1 | 12.51 | 39.17 | 19.06 | 3.01 |
| **d3-S04** | 1 | 10.06 | 20.52 | 19.30 | 3.06 |
| **d3-S05** | 1 | 15.17 | 16.56 | 19.02 | 3.43 |
| **d3-S06** | 1 | 17.18 | 34.71 | 19.02 | 3.43 |
| **d3-S07** | 2 | 10.70 | 271.64 | 21.90 | 0.72 |
| **d3-S08** | 2 | 10.70 | 211.89 | 21.90 | 0.72 |
| **d3-S09** | 2 | 8.44 | 9.66 | 21.76 | 0.69 |
| **d3-S10** | 3 | 11.10 | 40.68 | 23.34 | 0.33 |
| **d3-S11** | 4 | 9.57 | 71.25 | 23.84 | 0.27 |
| **d3-S12** | 4 | 10.42 | 186.28 | 23.92 | 0.27 |
| **d3-S13** | 5 | 8.01 | 26.89 | 24.25 | 0.25 |
| **d3-S14** | 5 | 8.22 | 62.21 | 24.25 | 0.25 |
| **d3-S15** | 5 | 11.09 | 82.43 | 24.25 | 0.25 |
| **d3-S16** | 5 | 10.45 | 103.45 | 24.25 | 0.25 |
| **d3-S17** | 5 | 9.00 | 61.13 | 24.22 | 0.25 |
| **d3-S18** | 5 | 10.91 | 46.16 | 24.32 | 0.20 |
| **d3-S19** | 6 | 13.80 | 47.63 | 24.71 | 0.24 |
| **d3-S20** | 9* | 11.78 | 310.61 | 24.95 | 0.25 |
| **d3-S21** | 9* | 10.38 | 97.36 | 24.95 | 0.25 |
| **d3-S22** | 9* | 11.02 | 102.61 | 24.95 | 0.25 |
| **d3-S23** | 9* | 15.83 | 93.82 | 24.95 | 0.25 |
| **d3-S24** | 9* | 9.98 | 320.66 | 25.03 | 0.24 |
| **d3-S25** | 9* | 14.22 | 210.84 | 25.03 | 0.24 |
| **d3-S26** | 9* | 18.59 | 167.32 | 25.03 | 0.24 |
| **d3-S27** | 9* | 21.13 | 227.11 | 25.03 | 0.24 |
| **d3-S28** | 9* | 20.71 | 201.36 | 25.03 | 0.24 |
| **d3-S29** | 9* | 20.61 | 189.17 | 25.03 | 0.24 |
| **d3-S30** | 9* | 18.38 | 172.28 | 25.03 | 0.24 |
| **d3-S31** | 9* | 13.44 | 153.08 | 25.03 | 0.24 |
| **d3-S32** | 9* | 10.56 | 299.66 | 25.03 | 0.24 |
| **d3-S33** | 9* | 18.68 | 268.68 | 25.03 | 0.24 |
| **d3-S34** | 9* | 20.95 | 208.03 | 25.03 | 0.24 |
| **d3-S35** | 9* | 21.01 | 249.02 | 25.03 | 0.24 |
| **d3-S36** | 9* | 19.12 | 224.80 | 25.03 | 0.24 |
| **d3-S37** | 9* | 18.18 | 150.41 | 25.03 | 0.24 |
| **d3-S38** | 9* | 19.91 | 159.42 | 25.03 | 0.24 |
| **d3-S39** | 9* | 21.06 | 182.49 | 25.03 | 0.24 |
| **d3-S40** | 9* | 18.74 | 178.32 | 25.03 | 0.24 |
| **d3-S41** | 9* | 16.60 | 140.04 | 25.03 | 0.24 |
| **d3-S42** | 9* | 20.07 | 149.57 | 25.03 | 0.24 |
| **d3-S43** | 9* | 21.39 | 217.19 | 25.03 | 0.24 |
| **d3-S44** | 9* | 15.33 | 103.03 | 25.03 | 0.24 |
| **d3-S45** | 9* | 15.79 | 126.47 | 25.03 | 0.24 |
| **d3-S46** | 9* | 18.67 | 146.57 | 25.03 | 0.24 |
| **d3-S47** | 9* | 19.54 | 156.13 | 25.03 | 0.24 |
| **d3-S48** | 9* | 15.20 | 101.92 | 25.03 | 0.24 |
| **d3-S49** | 9* | 10.36 | 65.06 | 25.03 | 0.24 |
| **d3-S50** | 11* | 16.03 | 445.53 | 25.07 | 0.25 |
| **d3-S51** | 11* | 12.08 | 48.86 | 25.07 | 0.25 |
| **d3-S52** | 11* | 16.22 | 393.13 | 25.07 | 0.25 |
| **d3-S53** | 11* | 15.61 | 96.70 | 25.07 | 0.25 |
| **d3-S54** | 26* | 18.90 | 267.28 | 25.09 | 0.21 |
| **d3-S55** | 95* | 64.42 | 145.30 | 25.17 | 0.08 |
| **d3-S56** | 95* | 64.42 | 144.93 | 25.17 | 0.08 |
| **d3-S57** | 95* | 64.52 | 73.39 | 25.17 | 0.08 |
| **d3-S58** | 95* | 59.07 | 81.39 | 25.17 | 0.08 |
| **d3-S59** | 95* | 58.97 | 80.82 | 25.17 | 0.08 |
| **d3-S60** | 95* | 59.56 | 82.69 | 25.17 | 0.08 |

**Table S2.** Complete dataset measured for $MoS_2$ resonators 2 μm in diameter. Note that the values marked with (*) have been determined only by AFM (see Raman and Photoluminescence Characterization section) and thus they present an uncertainty of about 5%.

| Sample | # layers | Frequency (MHz) | Quality factor | $A_{1g} – E^1_{2g}$ ($cm^{-1}$) | Norm. PL intensity |
|---|---|---|---|---|---|
| **d2-S01** | 1 | 30.82 | 42.14 | 19.30 | 3.06 |
| **d2-S02** | 1 | 19.10 | 102.25 | 19.02 | 3.43 |
| **d2-S03** | 1 | 21.33 | 105.29 | 19.02 | 3.43 |
| **d2-S04** | 2 | 10.46 | 234.96 | 21.90 | 0.72 |
| **d2-S05** | 2 | 10.28 | 196.39 | 21.90 | 0.72 |
| **d2-S06** | 2 | 14.47 | 31.70 | 21.81 | 0.74 |
| **d2-S07** | 2 | 11.44 | 26.04 | 21.66 | 0.66 |
| **d2-S08** | 2 | 17.29 | 119.82 | 21.66 | 0.66 |
| **d2-S09** | 2 | 16.42 | 17.15 | 21.84 | 0.69 |
| **d2-S10** | 2 | 13.57 | 32.06 | 21.66 | 0.66 |
| **d2-S11** | 3 | 11.61 | 363.60 | 23.18 | 0.34 |
| **d2-S12** | 4 | 10.36 | 183.12 | 23.88 | 0.29 |
| **d2-S13** | 4 | 18.04 | 129.20 | 23.84 | 0.27 |
| **d2-S14** | 5 | 12.52 | 289.15 | 24.25 | 0.25 |
| **d2-S15** | 5 | 12.11 | 210.00 | 24.25 | 0.25 |
| **d2-S16** | 5 | 11.90 | 257.34 | 24.25 | 0.25 |
| **d2-S17** | 5 | 11.57 | 297.28 | 24.22 | 0.25 |
| **d2-S18** | 6 | 19.38 | 27.19 | 24.71 | 0.24 |
| **d2-S19** | 9* | 28.83 | 109.26 | 24.95 | 0.25 |
| **d2-S20** | 9* | 24.82 | 182.48 | 24.95 | 0.25 |
| **d2-S21** | 9* | 26.99 | 221.66 | 24.95 | 0.25 |
| **d2-S22** | 9* | 25.65 | 177.18 | 24.95 | 0.25 |
| **d2-S23** | 9* | 24.51 | 195.86 | 25.03 | 0.24 |
| **d2-S24** | 9* | 29.32 | 240.77 | 25.03 | 0.24 |
| **d2-S25** | 9* | 33.80 | 206.22 | 25.03 | 0.24 |
| **d2-S26** | 9* | 36.15 | 262.50 | 25.03 | 0.24 |
| **d2-S27** | 9* | 35.35 | 233.12 | 25.03 | 0.24 |
| **d2-S28** | 9* | 33.97 | 259.78 | 25.03 | 0.24 |
| **d2-S29** | 9* | 30.56 | 191.63 | 25.03 | 0.24 |
| **d2-S30** | 9* | 23.73 | 127.64 | 25.03 | 0.24 |
| **d2-S31** | 9* | 27.55 | 205.22 | 25.03 | 0.24 |
| **d2-S32** | 9* | 34.87 | 228.56 | 25.03 | 0.24 |
| **d2-S33** | 9* | 35.09 | 238.36 | 25.03 | 0.24 |
| **d2-S34** | 9* | 34.14 | 230.10 | 25.03 | 0.24 |

## Supp. Info references